\documentclass{ws-ijmpa}

\usepackage[super,compress]{cite}
\usepackage{graphicx}
\begin{document}
	\markboth{Authors' Names}{Instructions for typing manuscripts (paper's title)}
	
	\catchline{}{}{}{}{}
	%
	%%%%%%%%%%%%%%%%%%%%%%%%%%%%%%%%%%%%%%%%%%%%%%%%%%%%%%%%%%%%%%%%%%%%
	
	\title{Masses of higher excited states of $\Xi_{c}'$ and $\Omega_{c}$ baryons
		%\footnote{For the title, try not to use more than 3 lines. Typeset the title in 10 pt roman, uppercase and boldface.}
	}
	
	\author{Pooja Jakhad$^{*}$, Juhi Oudichhya$^{\dagger}$ and Ajay Kumar Rai$^{\ddagger}$
		%\footnote{Typeset names in 8 pt roman, uppercase. Use the footnote to indicate the present or permanent address of the author.}
	}
	
	\address{Depertment of physics\\Sardar Vallabhbhai National Institute of Technology \\
		Surat 395007, Gujarat, India 
		%\footnote{State completely without abbreviations, the affiliation and mailing address, including country. Typeset in 8 pt italic.}
		\\
		*poojajakhad6@gmail.com
		\\
		$\dagger$juhioudichhya01234@gmail.com
		\\
		$\ddagger$raiajayk@gmail.com}

	\maketitle
	
	\begin{history}
		\received{Day Month Year}
		\revised{Day Month Year}
	\end{history}
	
	\begin{abstract}
		
		%	Excited states masses including $1F$-wave and $1G$-wave of  $\Xi_{c}'$ and $\Omega_{c}$ baryons  are calculated using the Relativistic Flux tube Model.
		%	Chiral effective theory of scalar and vector diquarks is formulated according to the linear sigma model.
		%	The main application is to describe the ground and excited states of singly heavy baryons with a charm or
		%	bottom quark. Applying the potential quark model between the diquark and the heavy quark (Q ¼ c, b), we
		%	construct a heavy-quark–diquark model. The spectra of the positive- and negative-parity states of  are obtained. The masses and interaction parameters of the effective theory are fixed partly
		%	from the lattice QCD data and also from fitting low-lying heavy-baryon masses. We find that the negative
		%	parity excited states of  (flavor 3) are different from those of , because of the inverse hierarchy of the
		%	pseudoscalar diquark. On the other hand,  (flavor 6) baryons have similar spectra. We
		%	compare our results of the heavy-quark–diquark model with experimental data as well as the q
	\end{abstract}
	Significant efforts have been made in recent years to measure the properties of charmed baryons. In this work, we study the $\Xi_{c}'$ and $\Omega_{c}$ baryons in the relativistic flux tube model with a quark-diquark picture of a baryon. The modified Regge relation between mass and angular momentum is used to predict the spin-average masses. The spin-dependent interactions are also included to compute the spin-dependent splitting. We calculate the masses of states belonging to the $1F$-wave and $1G$-wave of $\Xi_{c}'$ and $\Omega_{c}$ baryons, which are not yet detected experimentally. Our mass predictions can offer valuable insights to experimental facilities in their search to discover the higher excited states of $\Xi_{c}'$ and $\Omega_{c}$ baryons.
	\ccode{PACS numbers:}
	
	%\tableofcontents
	
	\section{Introduction}	
	
	\begin{figure*}
		\hspace{-0.65 cm}\includegraphics[scale=0.45]{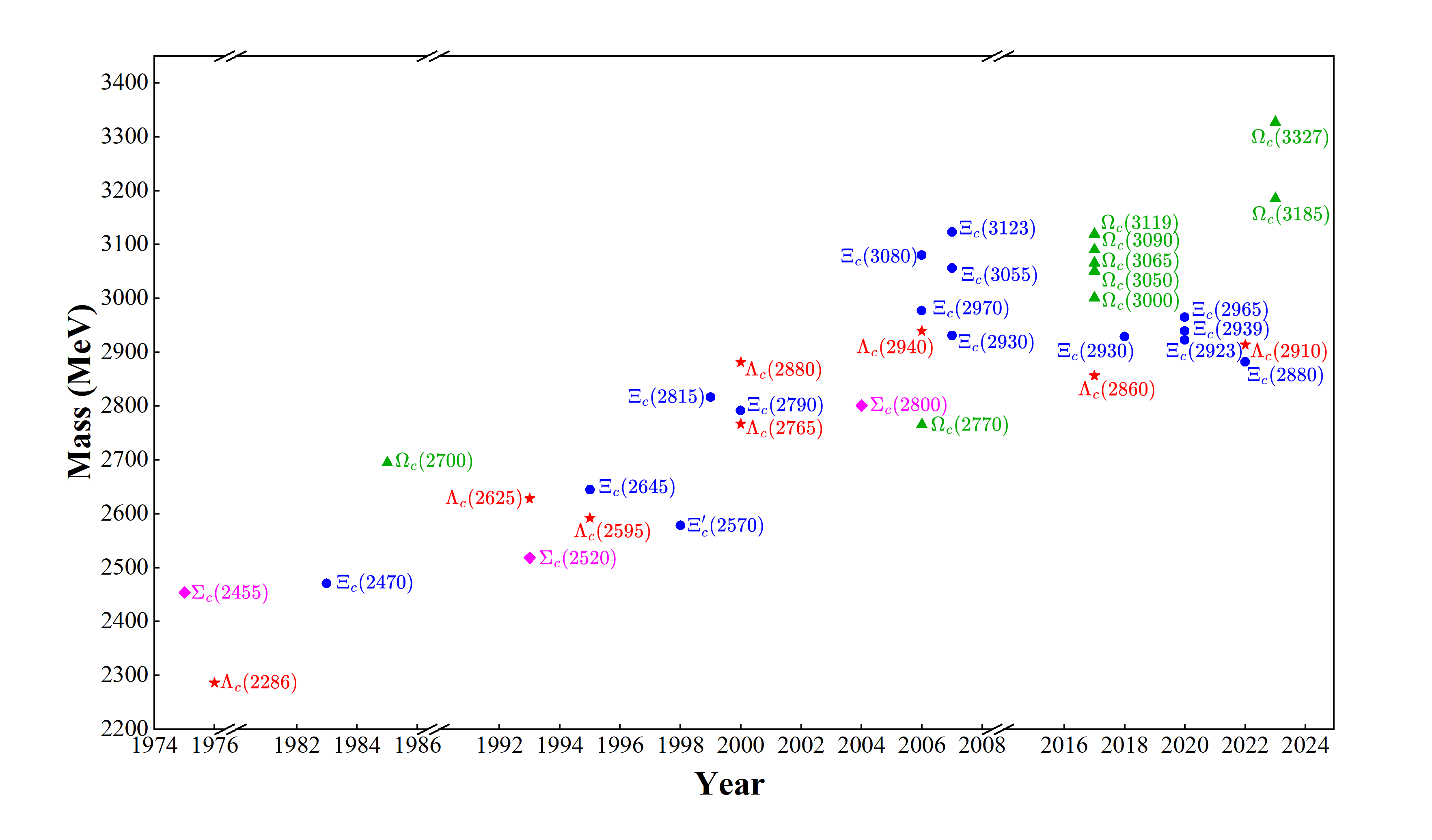}
		\caption{\label{fig:1} The experimentally observed singly charmed baryons \cite{PDG, belle2023, omegac3185and3327, lhc2023  }    }
	\end{figure*}
	Singly charmed baryons, consisting of one heavy charm quark and two light quarks, offer an ideal system for investigating the ideas of heavy quark symmetry and chiral symmetry. The study of charmed baryon spectroscopy, both through experimental and theoretical approaches, has made rapid progress in recent years. A careful investigation can enhance our understanding of the behavior of light quarks in the vicinity of a heavy charm quark.
	
	Experiments have observed several states of singly charmed baryons, as displayed in Fig. \ref{fig:1}.
	The $\Lambda_{c}$ baryonic family is the most widely recognized among the singly charmed baryons, with firmly established states in the $1S$-, $2S$-, $1P$-, and $1D$-waves. While for the $\Sigma_{c}$ baryonic family, only states in the $1S$-wave configuration with $J^{P}=1/2^{+}$ and $3/2^{+}$ have been identified. The only observed excited state of the $\Sigma_{c}$ baryon is the $\Sigma_{c}(2800)$, which has been detected by the Belle and \textit{BABAR} collaborations in the $\Lambda_{c}^{+}\pi$ channel \cite{Sigma_c2005,Sigma_c2008}. The spin-parity of this state is still unknown. 
	The $ \Xi_{c} $ baryonic family includes well-established states in the $1S$- and $1P$-waves. However, only the $1S$-wave states for the $\Xi_{c}' $ baryonic family have been identified. The excited states of $ \Xi_{c} $ and $ \Xi_{c}' $ baryons are hard to distinguish experimentally. Therefore, Partical Data Group (PDG) uses the notation $\Xi_{c} $ to represent the experimentally observed excited states of both $\Xi_{c} $ and $\Xi_{c}' $ baryons. Apart from the experimentally established states of $ \Xi_{c} $ and $ \Xi_{c}' $ baryons, several excited states, named $ \Xi_{c}(2880)$, $ \Xi_{c}(2923) $, $ \Xi_{c}(2930)$, $ \Xi_{c}(2939)$, $ \Xi_{c}(2965) $, $\Xi_{c}(2970) $, $ \Xi_{c}(3055)$, $ \Xi_{c}(3080)$, and $ \Xi_{c}(3123)$, have been observed by different experimental facilities. The spin-parity of these states, except $\Xi_{c}(2970) $, is not known yet. The discovery of the $ \Xi_{c}(3080)$ state was initially documented by Belle \cite{Cas_c2970_1} in 2006. After 2 years of this discovery, in 2008, this state was also observed by the \textit{BABAR} Collaboration \cite{BABARCac_c}. In the same year, the \textit{BABAR} Collaboration first detected the $ \Xi_{c}(3055)^{+} $ and $ \Xi_{c}(3123)^{+} $ states in the $ \Sigma_{c}(2455)^{++}K^{-} $ and $ \Sigma_{c}(2520)^{++}K^{-} $ channels, respectively \cite{BABARCac_c}. Belle also verified the $ \Xi_{c}(3055)^{+} $ state in 2014; however, they didn't find any signal for the $ \Xi_{c}(3123)^{+} $ state \cite{BelleCAs_c}. In 2020, the LHCb experiment came up with the discovery of the three excited $\Xi_{c}^{0}$ resonances, specifically named $\Xi_{c}(2923)^{0}$, $\Xi_{c}(2939)^{0}$, and $\Xi_{c}(2965)^{0}$, in the mass spectrum of $\Lambda_{c}^{+}K^{-}$ \cite{LHCb2020Cas_c}. Then in 2023, LHCb confirmed that the previously observed broad peak by \textit{BABAR}  \cite{BABARCac_c1} and Belle \cite{EPJCCas_c,EPJCCas_c1} for the $ \Xi_{c}(2930)^{0} $ state is resolved into two distinct peaks corresponding to $\Xi_{c}(2923)^{0}$ and $\Xi_{c}(2939)^{0}$ states \cite{lhc2023}. Moreover, the mass of the $ \Xi_{c}(2965)^{0} $ state is very close to the previously observed $ \Xi_{c}(2970)^{0} $ state \cite{Cas_c2970_1,Cas_c2970_2,BABARCac_c}. Hence, additional investigation is necessary to determine the equivalence of the states $ \Xi_{c}(2965)^{0} $ and $ \Xi_{c}(2970)^{0} $. Very recently, in 2023, the LHCb also obtained the very first evidence of a new $\Xi_{c}(2880)^{0}$ state \cite{lhc2023}. For $\Omega_{c}$ baryonic family only states belonging to $1S$-wave, named $ \Omega_{c}^{0} $ and $ \Omega_{c}(2770)^{0} $, is established. In 2017, the LHCb made a great announcement about the first observation of the five narrow excited states of $\Omega_{c}^{0} $ baryon, namely $ \Omega_{c}(3000)^{0} $, $ \Omega_{c}(3050)^{0} $, $ \Omega_{c}(3065)^{0} $, $ \Omega_{c}(3090)^{0} $, and $ \Omega_{c}(3120)^{0} $, in the $\Xi_{c}^{+}K^{-} $ channel \cite{LHCbOmega_c}. Later, except for $ \Omega_{c}(3120)^{0} $, the other four states were confirmed by Belle collaboration \cite{BelleOmega_c}. Recently, in 2023, observation of two new excited states, $\Omega_{c}(3185)^{0}$  and $\Omega_{c}(3327)^{0}$  in the $\Xi_{c}^{+}K^{-}$ invariant-mass spectrum, was also revealed by the LHCb collaboration \cite{omegac3185and3327}.
	
	The recent experimental progress in charmed baryon spectroscopy has motivated various theoretical approaches such as quark model \cite{ebert2011,shah2016,chen2017,Gandhi:2019bju,luo2023,li2023,yu2023}, Regge phenomenology \cite{oudichhya2021,juhicharm}, heavy hadron chiral perturbation theory  \cite{cheng2015,kawakami2019}, lattice QCD \cite{perezrubio2015}, QCD sum rules \cite{zwang2011, mao2017}, and relativistic flux tube  model \cite{ dwang2011, chen2015, PhysRevD.101.034016}. 
	
	In our previous study\cite{poojaprd2023}, we calculated the mass spectra of $\Xi_{c}'$ and $\Omega_{c}$ baryons up to the $1D$-wave using a relativistic flux tube model. Recent experimental advancements motivate us to predict masses of even more excited states. Thus, in the present work, we focus on calculating the masses of the states belonging to the $1F$-wave and $1G$-wave, which have not been detected experimentally. For simplicity, we will omit the isospin splitting in the following sections.

	%STATES WHICH ARE NOT INCLUDED IN PDG {
		%citation for $\Lambda_{c}(2910)$ \cite{belle2023}
		%citation for $\Omega_{c}(3185)^{0}$  and $\Omega_{c}(3327)^{0}$ \cite{omegac3185and3327}.
		%citation for $\Xi_{c}(2880)$ \cite{lhc2023}
		%}

	The paper is structured as follows: In Sec. 2, we present a theoretical framework that describes the relativistic flux tube model and the spin-dependent interactions that were used in this study. In Sec. 3, we tabulate the obtained results and discuss their importance. Lastly, in Sec. 4, we present our conclusion.
	
	\section{Theoretical Framework}
	$\Xi_{c}'$ and $\Omega_{c}$ baryons are composed of one charm quark($c$) and two light quarks ($u$ or $d$  or $s$). According to the heavy quark symmetry,  coupling between the heavy charm quark and the light quark ($u$, $d$, or $s$) is weak\cite{isgur1991}. This suggests that two light quarks can  correlate first to form a diquark within $\Xi_{c}'$ and $\Omega_{c}$ baryons. This picture has been supported by many theoretical studies \cite{Hooft, Semay2008, Narodetskii2009}. Hence, we study  of $\Xi_{c}'$ and $\Omega_{c}$ baryons in light-diquark-heavy-charm quark picture $(c-qq)$. 
	
	As a first step to obtain the mass spectra, we use the relativistic flux tube model \cite{PhysRevD.10.4262,PhysRevD.31.2910,lacourse1989,PhysRevD.45.4307,PhysRevD.48.417,PhysRevD.49.4675,Olsson1994,PhysRevD.51.3578,PhysRevD.53.4006,PhysRevD.60.074026} to calculate the spin average mass of $1F$-wave and $1G$-wave. In this model, two light quarks (qq)  correlate strongly  to form a diquark and then this diquark is connected  with a  charm quark through a  flux tube. The gluonic field which connects the diquark and a charm quark lies in this straight flux tube having string tension $\mathcal{T}$. The entire system consisting of charm quark, diquark, and flux tube is constantly revolving relativistically at $\omega $ angular speed around its center of mass. The Lagrangian of this rotating system is \cite{lacourse1989}
	\begin{equation}
		\label{eq:1}
		\mathcal{L}=\displaystyle\sum_{i=qq, c} \left[m_{i}(1-{v_{i}}^{2})^{\frac{1}{2}}+\frac{\mathcal{T}}{\omega}\int_{0}^{v_{i}}dv(1-v^{2})^{\frac{1}{2}}\right],
	\end{equation}
	where $m_{qq}$ and $m_{c}$ represents the current quark masses of diquark and charm quark, respectively. We opted to use the speed of light equal to one in natural units for convenience.  $v_{qq}$ and $v_{c}$ specifies the speed of diquark and charm quark, respectively.
	From this, one can derive the angular momentum  of the system as 
	\begin{equation}
		\label{eq:2}
		L=\frac{\partial \mathcal{L}}{\partial \omega}=\displaystyle\sum_{i=qq, c}  \left[\frac{m_{i}v_{i}^{2}}{\omega\sqrt{1-v_{i}^{2}}}+\frac{\mathcal{T}}{\omega^{2}}\int_{0}^{v_{i}}\frac{v^{2}dv}{(1-v)^{2}}\right].
	\end{equation}
	The mass of the $ cqq $ baryon ($M$) is equal to the Hamiltonian of the system ($ H $), which is  given by
	\begin{equation}
		\label{eq:3}
		H={M}=\displaystyle\sum_{i=qq, c}  \left[\frac{m_{i}}{\sqrt{1-v_{i}^{2}}}+\frac{\mathcal{T}}{\omega}\int_{0}^{v_{i}}\frac{dv}{(1-v)^{2}}\right].
	\end{equation}
	By integrating Eq.(\ref{eq:2}) and  Eq.(\ref{eq:3}), we obtain

	%Here, we represent mass by $\bar{M}$ to show that spin interactions are not included in this mass and we can consider it as a spin average mass.
	
	\begin{equation}
		\label{eq:4}
		L=\frac{1}{\omega}(M_{qq}v_{qq}^{2}+M_{c}v_{c}^{2})+\frac{\mathcal{T}}{2\omega^{2}}\displaystyle\sum_{i=qq, c} (sin^{-1}v_{i}-v_{i}\sqrt{1-v_{i}^2}),
	\end{equation}
	and
	\begin{equation}
		\label{eq:5}
		{M}= M_{qq}+M_{c}+\frac{T}{\omega}\displaystyle\sum_{i=qq, c} sin^{-1}v_{i},
	\end{equation}
	where   $ M_{qq}=m_{qq}/\sqrt{1-v_{qq}^{2}} $ and $ M_c=m_c/\sqrt{1-v_{c}^{2}} $ are the effective masses of  diquark and charm quark, respectively. \\
	The boundary condition at the charm quark end of the flux tube turns into\cite{chen2015}
	\begin{equation}
		\label{eq:6}
		\frac{\mathcal{T}}{\omega}=
		%\frac{M_c v_c}{1-v_{c}^{2}}=
		\frac{M_c v_c}{\sqrt{1-v_{c}^{2}}}=M_c v_c[1+\frac{v_c^2}{2}+\frac{3v_c^4}{8}+...]\simeq M_c v_c.
	\end{equation}
	As $ M_{c}\gg M_{qq}$ for singly charmed baryons, the $c$-quark moves non-relativistically in baryon, implying that $ v_{c} $ is much less than 1. Therefore, the higher order terms of $ v_{c} $ are omitted in above relation (\ref{eq:6}). In the case of a very small diquark mass, we assume that it move ultra-relativistically i.e.  $ v_{qq}\rightarrow 1 $. By expanding Eq. (\ref{eq:4}) and (\ref{eq:5}) in a power series of $ v_{c} $ up to the second term and applying Eq. (\ref{eq:6}), we obtain a relation between the mass and angular momentum of the system as \cite{poojaprd2023}
	\begin{equation}
		\label{eq:7}
		({M} -M_c)^2=\frac{\sigma}{2}L+(M_{qq}+M_c v_{c}^{2}),
	\end{equation}
	where $\sigma=2\pi \mathcal{T}$. Using the this relation, one may compute the spin average mass of $1F$-wave and $1G$-wave  by putting $L=3$ and $L=4$, respectively.\\
	In the relativistic flux tube model, the distance between a charm quark and a diquark is \cite{chen2015}
	\begin{equation}
		\label{eq:8}
		r=\frac{v_{qq}+v_c}{\omega}=(v_{qq}+v_c)\sqrt{\frac{8L}{\sigma}}.
	\end{equation}
	
	Further, as a second step, we incorporate the spin dependent interactions\cite{chen2022}. The contribution to mass resulting from spin-dependent interactions can be expressed as
	\begin{equation}
		{M_{SD}}= H_{so}+ H_{t}+ H_{ss},
	\end{equation} 
	where
	\begin{equation} 
		H_{so}=  [(\frac{2\alpha}{3r^3}-\frac{b}{2r}) \frac{1}{m_{qq}^2 }+\frac{4\alpha}{3r^3} \frac{1}{m_{qq} m_c }]\mathbf{L}.\mathbf{S_{qq}}
		+[(\frac{2\alpha}{3r^3}-\frac{b}{2r}) \frac{1}{m_c^2 }+\frac{4\alpha}{3r^3} \frac{1}{ m_{qq} m_c}]\mathbf{L}.\mathbf{S_c},
	\end{equation} 
	\begin{equation} 
		H_t=\frac{4\alpha}{3r^3} \frac{1}{m_{qq} m_c} [\frac{3(\mathbf{S_{qq}}.\mathbf{r})(\mathbf{S_c}.\mathbf{r})}{r^2} -\mathbf{S_{qq}}.\mathbf{S_c}],
	\end{equation} 
	and
	\begin{equation}
		H_{ss}=\frac{32\alpha\sigma_0^3}{9\sqrt{\pi} m_{qq} m_c}e^{-\sigma_0^2 r^2 } \mathbf{S_{qq}}.\mathbf{S_c}.
	\end{equation}
	Here, $H_{so}$, $H_{t}$ and $H_{ss}$ stands for spin-orbit interaction, tensor interaction and spin-spin interaction, respectively. $ \mathbf{S_c} $, $ \mathbf{S_{qq}} $, and $\mathbf{L}$ represent the spin operator of the charm quark, the spin operator of the diquark, and the orbital angular momentum operator of the system, respectively. $\alpha$, $b$, and $\sigma_0$ are the parameters that we calculate from the masses of the experimentally observed states.  As a result of these spin-dependent interactions ($M_{SD}$), the states with mass ${M}$ will undergo a splitting into distinct states with mass $M+M_{SD}$.

	As $\Xi_{c}'$ and $\Omega_{c}$ baryons belong to the flavour symmetric sextet, the Pauli's exclusion principle suggests that the spin state of diquark has to antisymmetric implying that $S_{qq}=1$. Thus, one possibility is that $\mathbf{S_{qq}}$ couples with $\mathbf{S_{c}}$ to form the total spin angular momentum $\mathbf{S}$, and the next $\mathbf{S}$ couples with $\mathbf{L}$ to form $\mathbf{J}$. This is called the L-S coupling scheme. However, as $ M_{c} $ is much greater than $M_{qq}$ for $\Xi_{c}'$ and $\Omega_{c}$ baryons, the preferred coupling scheme is the j-j coupling scheme. In this scheme, the coupling between the spin of the diquark ($\mathbf{S_{qq}}$) and the orbital angular momentum ($\mathbf{L}$) results in the total angular momentum of the diquark ($\mathbf{j}$), which is then coupled with the spin of the charm quark ($\mathbf{S_{c}}$) to form the total angular momentum of the system ($\mathbf{J}$). Hence, we refer to baryonic states as j-j coupling states, which are defined by $j$ and $J$, as shown in Table. \ref{tab:table2} and \ref{tab:table3}.

	\section{Results and Discussions}
	
	In the above theoretical framework, we calculate the masses of the states belonging to the $1F$- and $1G$-waves of $\Xi_{c}'$ and $\Omega_{c}$ baryons using the parameters that are extracted from the well-established states of singly charmed baryons as listed in Table \ref{tab:table1}. The calculated results for $\Xi_{c}'$ and $\Omega_{c}$ baryons are listed in Tables \ref{tab:table2} and \ref{tab:table3}, respectively. The very first column of this table displays the baryonic states represented as $\textit{$|nL, J^P\rangle_{j}$}$ in the j-j coupling scheme,  accompanied by our calculated masses in the framework of the relativistic flux tube model in the second column. Further, as the experimental data is not available for $1F$- and $1G$-wave states, we compare our calculated masses with those of the other theoretical models in the remaining columns. In a study by Ebert et al.\cite{ebert2011}, a relativistic quark potential model (RQPM) is used to look at heavy baryon mass spectra up to high levels of radial and orbital excitation. In ref. \cite{oudichhya2021,juhicharm}, the Regge phenomenology (RP) has been employed to study the mass spectra of singly charmed baryons. Further in the recent article, Luo et al.\cite{luo2023} have investigated the spectroscopic properties of $1F$-wave singly charmed baryons in a non-relativistic potential model (NRPM). In ref.\cite{li2023,yu2023}, the Godfrey-Isgur relativized quark model (GIRQM) is used to study singly heavy baryons. Further, the authors in ref.\cite{shah2016} used the hyper central constituent quark model (HCQM) to study the singly charmed baryons with three body picture.
	
	\begin{table}[h]
		\tbl{The values of parameters involved in our theorteical model \cite{poojaprd2023}.}
		{\begin{tabular}{@{}cccccccc@{}}
				\toprule
				Parameters         & $\sigma$ (GeV$^{2}$) & $M_{qq}$ (GeV) & $M_{c} (GeV)$ & $v_{c}$ & $\alpha$ & $b$ (GeV$^{2}$) & $\sigma_{0}$ (GeV) \\
				\colrule
				$\Xi_{c}'$         &  1.856 &  0.841 & 1.448 &  0.48 & 0.426 & -0.076  & 0.400      \\
				$\Omega_{c}$       &  2.026 &  0.959 & 1.448 &  0.48 & 0.426 & -0.076  & 0.425      \\
				\botrule          
			\end{tabular} \label{tab:table1}}
	\end{table}

	\begin{table}[h]
		\tbl{Theoretical masses of $\Xi_{c}'$ baryons (in MeV) calculated using the relativistic flux tube model, the relativistic quark potential model (RQPM)\cite{ebert2011},  the Regge phenomenology (RP) \cite{juhicharm}, the  Godfrey-Isgur relativized quark model (GIRQM)\cite{li2023}, and  the non-relativistic quark potential model (NQPM)  \cite{luo2023}.}
		{\begin{tabular}{@{}cccccc@{}}
				\toprule
				States                  & This work & RQPM\cite{ebert2011} & RP\cite{juhicharm} &GIRQM\cite{li2023}  & NQPM\cite{luo2023} \\
				\textit{$|nL, J^P\rangle_{j}$ }   &         &                  &	               &		     	& 			     \\
				\colrule
				$|1F, 3/2^-\rangle_{j=2}$     & 3409.0  &       3418       &                  &	3424	    & 	3427		 \\
				$|1F, 5/2^-\rangle _{j=2}$     & 3437.4  &       3394       &       3336       &	3428 		&	3433	     \\
				$|1F, 5/2^-\rangle_{j=3}$     & 3458.9  &       3408       &       3402       &	3424 		& 	3408		 \\
				$|1F, 7/2^-\rangle _{j=3}$     & 3524.1  &       3393       &       3461       &	3428		& 	3412         \\
				$|1F, 7/2^-\rangle_{j=4}$     & 3484.6  &       3373       &       3279       & 	3423 		& 	3382	     \\
				$|1F, 9/2^-\rangle_{j=4}$     & 3547.3  &       3357       &       3541       &	3428		& 	3383	     \\
				\noalign{\smallskip}			
				$|1G, 5/2^+\rangle_{j=3}$     & 3627.8  &       3623       &                  &	 	    	& 	             \\
				$|1G, 7/2^+\rangle _{j=3}$     & 3654.4  &       3584       &       3520       &	    		& 	             \\
				$|1G, 7/2^+\rangle_{j=4}$     & 3679.3  &       3608       &       3646       &				& 	             \\
				$|1G, 9/2^+\rangle _{j=4}$     & 3742.9  &       3582       &       3709       &		 		& 	             \\
				$|1G, 9/2^+\rangle_{j=5}$     & 3703.9  &       3558       &       3448       &		 		& 			     \\
				$|1G, 11/2^+\rangle_{j=5}$     & 3765.8  &       3536       &       3792       &              &			     \\
				\botrule                 & 
			\end{tabular} \label{tab:table2}}
	\end{table}
	
	\begin{table}[h]
		\tbl{Theoretical masses of $\Omega_{c}$ baryons (in MeV) calculated using the relativistic flux tube model, the relativistic quark potential model (RQPM)\cite{ebert2011},  Godfrey-Isgur relativized quark model (GIRQM)\cite{yu2023},  the Regge phenomenology (RP) \cite{juhicharm}, the hyper-central constituent quark model (HCQM) \cite{shah2016}, and  the non-relativistic quark potential model (NQPM)  \cite{luo2023}.}
		{\begin{tabular}{@{}ccccccc@{}}
				\toprule
				States                  & This work & RQPM\cite{ebert2011} & GIRQM\cite{yu2023}& RP\cite{juhicharm} & HCQM\cite{shah2016} &  NQPM\cite{luo2023} \\
				\textit{$|nL, J^P\rangle_{j}$ }   &         &                  &              &	                &  			      &	                \\
				\colrule
				$|1F, 3/2^-\rangle_{j=2}$     & 3547.6  &       3533       &     3525     &                  &      3643       &  	3540		 \\
				$|1F, 5/2^-\rangle _{j=2}$     & 3576.9  &       3515       &     3528     &       3474       &      3602       &  	3547	     \\
				$|1F, 5/2^-\rangle_{j=3}$     & 3590.5  &       3522       &     3525     &       3545       &      3613       &   	3532		 \\
				$|1F, 7/2^-\rangle _{j=3}$     & 3646.4  &       3514       &     3529     &       3590       &      3577       &   	3537	     \\
				$|1F, 7/2^-\rangle_{j=4}$     & 3617.1  &       3498       &     3524     &       3419       &      3565       &   	3521	     \\
				$|1F, 9/2^-\rangle_{j=4}$     & 3670.5  &       3485       &     3529     &       3672       &      3532       &   	3520	     \\
				\noalign{\smallskip}
				$|1G, 5/2^+\rangle_{j=3}$     & 3772.7  &       3739       &     3719     &                  &                 &   			     \\
				$|1G, 7/2^+\rangle _{j=3}$     & 3800.3  &       3707       &     3720     &       3659       &                 &   			     \\
				$|1G, 7/2^+\rangle_{j=4}$     & 3816.4  &       3721       &     3719     &       3791       &                 &   		         \\
				$|1G, 9/2^+\rangle _{j=4}$     & 3870.3  &       3705       &     3720     &       3841       &                 &   			     \\
				$|1G, 9/2^+\rangle_{j=5}$     & 3841.9  &       3685       &     3719     &       3589       &                 &   			     \\
				$|1G, 11/2^+\rangle_{j=5}$     & 3894.1  &       3665       &     3720     &       3926       &                 &   			     \\
				\botrule                
			\end{tabular} \label{tab:table3}}
	\end{table}

	We observe that our calculated masses fall close to the masses predicted by these  theoretical models as shown in Table \ref{tab:table2} and \ref{tab:table3}. However, we also observe various model-dependent differences between the masses predicted by different theoretical models. For instance, the mass splitting within the $1F$- and $1G$-wave states of $\Xi_{c}'$ and $\Omega_{c}$ baryons is significantly smaller in the mass spectra predicted by GIRQM \cite{li2023, yu2023} and NQPM \cite{luo2023} compared to the mass spectra predicted by our model and other theoretical models listed in Table \ref{tab:table2} and \ref{tab:table3}. Even the GIRQM \cite{li2023, yu2023} predicts the same mass for two or three states of the $1F$- and $1G$-waves, while our model does not exhibit this inconsistency. Furthermore, in our predicted mass spectra, we observe that for spin doublet states, (states having identical $j$), the mass of the state with higher $J$ is greater than that of the state with lower $J$. This feature is also reflected in the mass spectra predicted by RP \cite{juhicharm}. Additionally, within our predictions, as the $j$ value increases for a given multiplet, the mass also increases. However, these features are notably absent in the mass spectra predicted by RQPM \cite{ebert2011} and HCQM \cite{shah2016}. In these models, the mass of the state with higher $J$ is actually less than that of the state with lower $J$ for spin doublet states, and for a given multiplet, the mass decreases as the $j$ value increases. These distinct characteristics observed in mass spectra from various theoretical models indicate the necessity for additional experimental measurements to effectively differentiate between them.

	In our earlier work, the masses of the experimentally reported states of $\Xi_{c}'$ and $\Omega_{c}$ baryons, such as $\Xi_{c}'$, $\Xi_{c}(2645)$, $\Xi_{c}(2880)$, $\Xi_{c}(2923)$, $\Xi_{c}(2939)$, $ \Xi_{c}(3123)$, $\Xi_{c}(3055)$, $\Omega_{c}$, $\Omega_{c}(2770)$, $ \Omega_{c}(3000)$, $ \Omega_{c}(3050)$, $ \Omega_{c}(3065)$, $ \Omega_{c}(3090)$, and $ \Omega_{c}(3120)$, were successfully reproduced using a relativistic flux tube model with a quark-diquark picture of the baryon\cite{poojaprd2023}. In that work, these states observed in the experiment were predicted to be the states of $1S$-, $2S$-, $1P$-, and $1D$-waves. With the advancement of experimental facilities like LHCb, the study of singly charmed baryons is progressing rapidly, and the higher excited states, such as states belonging to the $1F$- and $1G$-waves, are expected to be observed. So our results about the masses of unobserved states of the $1F$- and $1G$-waves of $\Xi_{c}'$ and $\Omega_{c}$ baryons can help future experiments determine the missing resonances in different decay channels.

	\section{Conclusion}
	This study involves the determination of the masses of states associated with the $1F$- and $1G$-wave of $\Xi_{c}'$ and $\Omega_{c}$ baryons. The linear Regge relation is obtained from the relativistic flux tube model, assuming that the vector diquark in $\Xi_{c}'$ and $\Omega_{c}$ baryons remains in its ground state. The spin dependent interactions are also included in the j-j coupling scheme.  Our findings about the unobserved higher excited states of $\Xi_{c}'$ and $\Omega_{c}$ baryons can provide valuable assistance in future experimental investigations conducted at LHCb, CMS, Belle II, BESIII \cite{BESIII2022}, and PANDA \cite{PANDA1,PANDA2,PANDA3,PANDA4,PANDA5}.

	\section*{Acknowledgments}
	The authors express gratitude to the organizers of the 11th international conference on new frontiers in physics (ICNFP 2022) for providing a platform to present our work.  Ms. Pooja Jakhad appreciates the financial support from the Council of Scientific \& Industrial Research (CSIR) through the JRF-FELLOWSHIP scheme, file no. 09/1007(13321)/2022-EMR-I.
	
	\appendix

	%	\section{Appendices}
	%	
	%	Appendices should be used only when absolutely necessary. They
	%	should come before the References. If there is more than one
	%	appendix, number them alphabetically. Number displayed equations
	%	occurring in the Appendix in this way, e.g.~(\ref{app1}),\break
	%	(\ref{app2}), etc.

	%\begin{thebibliography}{000} %for 3 digits
	%\begin{thebibliography}{00}  %for 2 digits
	
\end{document}